\documentclass[twocolumn,secnumarabic,amssymb, nobibnotes, aps, prl, floatfix, reprint,superscriptaddress, showpacs]{revtex4-1}
 \pdfoutput=1
\usepackage{graphicx}

\begin{document}
\bibliographystyle{apsrev4-1}

\title{Destabilization of magnetic order in a dilute Kitaev spin liquid candidate}%

\author{P. Lampen-Kelley}%
\affiliation{Department of Materials Science and Engineering, University of Tennessee, Knoxville, TN 37996, U.S.A.}
\affiliation{Materials Science and Technology Division, Oak Ridge National Laboratory, Oak Ridge, TN, 37831, U.S.A.}
\author{A. Banerjee}
\author{A.A. Aczel}
\author{H.B. Cao}
\affiliation{Quantum Condensed Matter Division, Oak Ridge National Laboratory, Oak Ridge, TN 37831, U.S.A.}
\author{M.B. Stone}
\affiliation{Quantum Condensed Matter Division, Oak Ridge National Laboratory, Oak Ridge, TN 37831, U.S.A.}
\author{C.A. Bridges}
\affiliation{Chemical Sciences Division, Oak Ridge National Laboratory, Oak Ridge, TN 37831, U.S.A.}
\author{J.-Q. Yan}
\affiliation{Materials Science and Technology Division, Oak Ridge National Laboratory, Oak Ridge, TN, 37831, U.S.A.}
\author{S.E. Nagler}
\affiliation{Quantum Condensed Matter Division, Oak Ridge National Laboratory, Oak Ridge, TN 37831, U.S.A.}
\author{D. Mandrus}
\affiliation{Department of Materials Science and Engineering, University of Tennessee, Knoxville, TN 37996, U.S.A.}
\affiliation{Materials Science and Technology Division, Oak Ridge National Laboratory, Oak Ridge, TN, 37831, U.S.A.}

\date{September 6, 2017}

\begin{abstract}
The insulating honeycomb magnet $\alpha$-RuCl$_{3}$ exhibits fractionalized excitiations that signal its proximity to a Kitaev quantum spin liquid (QSL) state, however, at $T=0$, fragile long-range magnetic order arises from non-Kitaev terms in the Hamiltonian. Spin vacancies in the form of Ir$^{3+}$ substituted for Ru are found to destabilize this long-range order. Neutron diffraction and bulk characterization of Ru$_{1-x}$Ir$_{x}$Cl$_{3}$ show that the magnetic ordering temperature is suppressed with increasing $x$ and evidence of zizag magnetic order is absent for $x>0.3$. Inelastic neutron scattering demonstrates that the signature of fractionalized excitations is maintained over the full range of $x$ investigated. The depleted lattice without magnetic order thus hosts a spin-liquid-like ground state that may indicate the relevance of Kitaev physics in the magnetically dilute limit of RuCl$_{3}$. 

\end{abstract}

\maketitle

The quantum spin liquid (QSL) holds particular fascination as a state of matter that exhibits strong quantum entanglement yet is devoid of  long-range order \cite{balents_spin_2010, savary_quantum_2017}. These exotic states can possess topologically protected fractionalized excitations, with possible implications for quantum information science \cite{nayak_non-abelian_2008, kitaev_fault-tolerant_2003}. A prototypcal example is the Kitaev model on a honeycomb lattice \cite{kitaev_anyons_2006}, which can be solved exactly and has a QSL ground state.  An effective Hamiltonian with Kitaev terms consisting of  bond-directional Ising couplings may arise in spin-orbit assisted Mott insulators with $J_{eff} = 1/2$ moments in an edge-sharing octahedral environment \cite{jackeli_mott_2009}. A strong push for the experimental realization of quasi-2D honeycomb lattices showing  Kitaev physics initially focused on iridate materials with the chemical formula A$_{2}$IrO$_{3}$ \cite{singh_antiferromagnetic_2010, singh_relevance_2012, hwan_chun_direct_2015, chaloupka_magnetic_2016}, and more recently $\alpha$-RuCl$_{3}$  \cite{plumb_$ensuremathalpha-mathrmrucl_3$:_2014, kim_kitaev_2015, banerjee_proximate_2016, kubota_successive_2015}. Each of these compounds orders magnetically at low temperatures in a zigzag or incommensurate phase \cite{liu_long-range_2011, ye_direct_2012, choi_spin_2012, williams_incommensurate_2016, sears_magnetic_2015, johnson_monoclinic_2015, cao_low-temperature_2016}, and the effective low energy Hamiltonian is believed to be described by a generalized Heisenberg-Kitaev-$\Gamma$ model \cite{rau_generic_2014, reuther_spiral_2014,chaloupka_zigzag_2013, chaloupka_hidden_2015,chaloupka_kitaev-heisenberg_2010, sizyuk_importance_2014, sizyuk_selection_2016, yadav_kitaev_2016, katukuri_kitaev_2014, foyevtsova_textitab_2013, winter_challenges_2016, shinjo_density-matrix_2015}. Despite the appearance of long-range order, broad scattering continua observed via inelastic neutron  or Raman scattering  in $\alpha$-RuCl$_{3}$ \cite{banerjee_proximate_2016,banerjee_neutron_2017,sandilands_scattering_2015, nasu_fermionic_2016}  and the iridates \cite{ nath_gupta_raman_2016} match the predicted signatures of itinerant Majorana fermions in pure Kitaev calculations \cite{knolle_dynamics_2014, knolle_raman_2014, perreault_resonant_2016}, suggesting that these materials are proximate to the QSL state and that Kitaev interactions play an important role.   

In this letter, we report the evolution of the magnetic ground state in $\alpha$-RuCl$_{3}$ with magnetic Ru$^{3+}$ substituted by nonmagnetic Ir$^{3+}$, and determine a phase diagram as a function of temperature and dilution. The motivation is two-fold: to understand the role of defects in Kitaev-candidate materials and to explore avenues towards suppression of long-range order. Numerous theoretical studies predict the emergence of novel superconductivity with hole doping in the strong Kitaev limit \cite{hyart_competition_2012, you_doping_2012, mei_possible_2012, okamoto_doped_2013, okamoto_global_2013, trousselet_hole_2014, kimme_symmetry-protected_2015} while bond disorder \cite{zschocke_physical_2015}, dislocations \cite{brennan_lattice_2016}, magnetic impurities \cite{dhochak_magnetic_2010}, and spin vacancies \cite{ willans_disorder_2010, willans_site_2011, g._localized_2012, andrade_magnetism_2014, sreejith_vacancies_2016, trousselet_effects_2011, halasz_doping_2014, halasz_coherent_2016} have also received theoretical attention. Experimentally, substitution of both magnetic and non-magnetic cations for Ir rapidly led to spin glass freezing in Na$_{2}$IrO$_{3}$ and Li$_{2}$IrO$_{3}$ \cite{mehlawat_fragile_2015, manni_effect_2014-1}. Isoelectronic substitution within the solid solution (Na,Li)$_{2}$IrO$_{3}$ decreased the magnetic ordering temperature, although phase separation has hampered efforts to completely suppress long-range order \cite{cao_evolution_2013, manni_effect_2014, rolfs_spiral_2015}.

The chemically simpler binary compound $\alpha$-RuCl$_{3}$ provides an excellent framework to explore the effects of various perturbations on the relevant physics.  Low-spin 5d$^{6}$ Ir$^{3+} (S = 0)$ represents a non-magnetic impurity in the $J_{\mathrm{eff}} = 1/2$ Ru$^{3+}$ magnetic sublattice while the identical ionic radii (0.68 \AA) of Ru$^{3+}$ and Ir$^{3+}$ preserve a regular MCl$_{6}$ environment (M = Ru, Ir).  

The van der Waals bonded honeycomb layers in $\alpha$-RuCl$_{3}$  are susceptible to stacking faults that are known to affect the magnetic ordering properties.  Single crystals without stacking faults show zigzag order with $T_{N}\simeq7$ K and a three-fold (ABC) out- of-plane periodicity. With stacking faults, a phase with two-fold periodicity (ABAB) also appears at $T_{N}\simeq14$~K \cite{banerjee_proximate_2016, cao_low-temperature_2016,johnson_monoclinic_2015,kubota_successive_2015}. Polycrystalline material shows only the 14~K transition. Our present study of Ru$_{1-x}$Ir$_{x}$Cl$_{3}$ reveals the suppression of the ABC and ABAB magnetic phases with critical concentrations of $x \simeq 0.1$ and $x \simeq 0.3$, respectively, demonstrating that site dilution represents a viable approach to destabilizing long-range magnetic order. Spectroscopic examination of the magnetically disordered limit reveals a dynamic ground state with indications of persistent fractionalization of spin excitations.

\begin{figure}[t]
\centering
\includegraphics[scale=0.28]{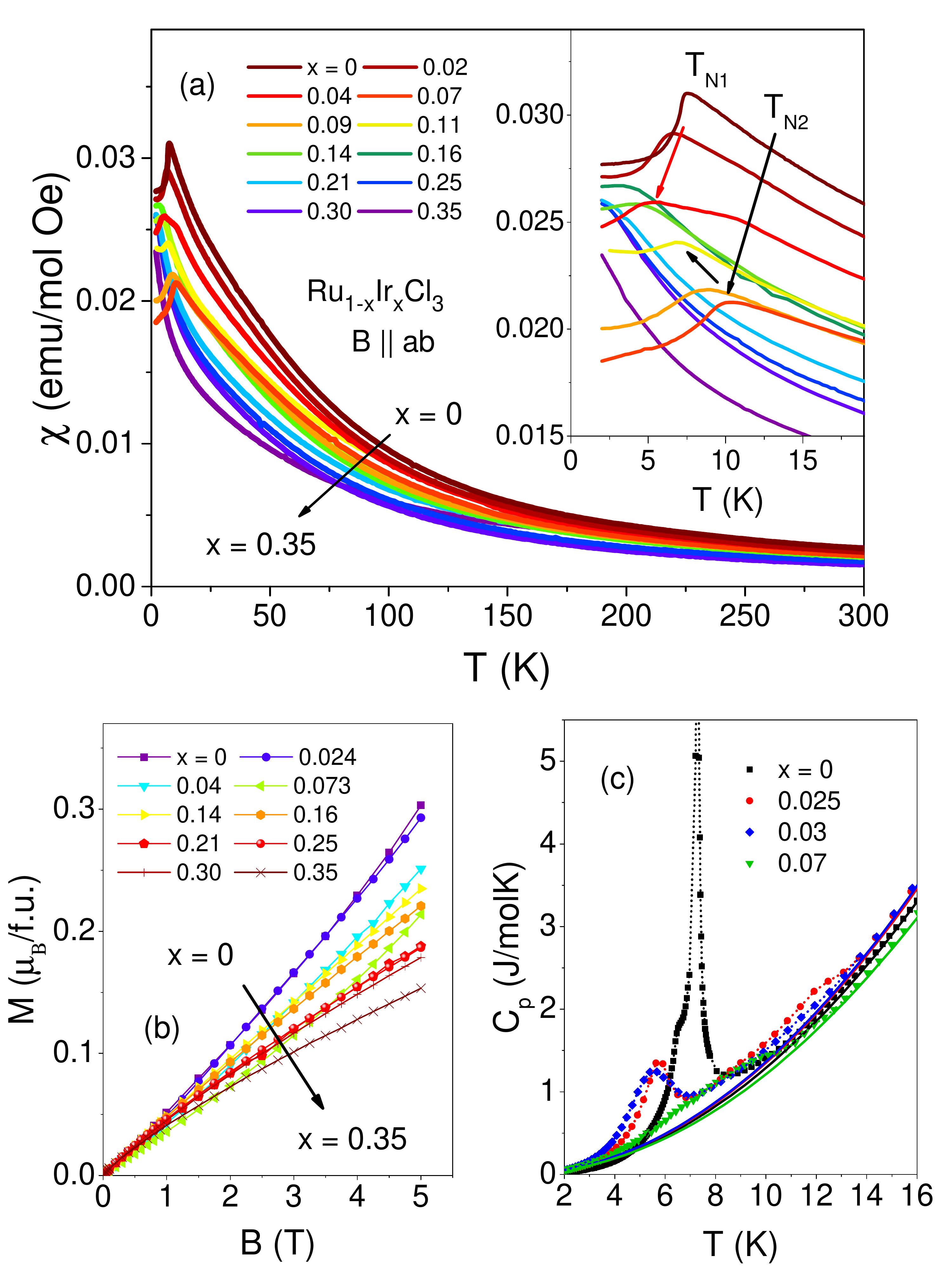}
\caption{(a) Magnetic susceptibility curves of Ru$_{1-x}$Ir$_{x}$Cl$_{3}$ single crystals ($x$ values indicated in legend) with a magnetic field of $B =$ 1~T applied in the $ab$ plane. Inset: magnification of low-temperature range. Red arrow and black/grey arrows mark the evolution of $T_{N1}$ and $T_{N2}$, respectively. (b) Field-dependent magnetization at 2~K. (c) Heat capacity curves of crystals with small Ir concentration. Solid lines are an estimate of the lattice contribution. Dotted lines are a guide to the eye.} 
\end{figure}

Figure 1(a) shows the magnetic susceptibility of a series of Ru$_{1-x}$Ir$_{x}$Cl$_{3}$ single crystals with magnetic field applied in the honeycomb $ab$ plane. In addition to the $T_{N1} =7$~K zigzag transition of the parent compound, a second feature at $T_{N2} \leq 14$~K for $x>0$ (black arrow, Fig.1a Inset) indicates that substituted crystals obtained from the current growth process are not completely free of stacking faults. The sharp cusp in the susceptibility at $T_{N1}$ is rapidly shifted toward lower temperatures at non-zero $x$ (red arrow) denoting a suppression of the ordering temperatures with dilution. As $x$ increases, the minority phase contribution ($T_{N2}$) gains prominence and is the only transition evident above 2~K by $x = 0.09$.  Further Ir substitution leads to a rounding of the cusp at $T_{N2}$, which decreases continuously with $x$ and disappears near the percolation limit of the honeycomb lattice, $x \sim 0.3$ \cite{feng_percolation_2008}. Notably, field-cooled and zero-field-cooled measurements are identical for all values of $x$, in contrast to the spin-glass-like characteristics reported in the honeycomb iridates with non-magnetic substitution \cite{manni_effect_2014}. In fact, muon spin relaxation ($\mu$SR) experiments rule out a glassy state and point to fast dynamics on the $\mu$SR time scale ($10^{4}-10^{12}$ Hz) in the magnetically disordered limit [see Supplementary Information (SI) Fig. S4 and S7].

The low-temperature susceptibility does not decrease monotonically with $x$, but instead increases over the range $0.07 < x < 0.16$. This behavior might arise from uncompensated moments introduced by nonmagnetic impurities in the ordered antiferromagnetic ground state of the parent compound. We also note that vacancies in a site diluted Kitaev model are predicted to increase the local susceptibility \cite{willans_disorder_2010, willans_site_2011, trousselet_effects_2011}. In the parent compound the field-dependent magnetization at 2~K shows an upward curvature approaching the suppression of the zigzag phase near 7.5~T \cite{johnson_monoclinic_2015,kubota_successive_2015}. For $x >$ 0.09, the field-dependent magnetization curves in Fig. 1(b) develop opposite concavity from the clean limit. Interestingly, this coincides with the region of increase in the low temperature susceptibility mentioned above. 

The downward trend in $T_{N1}$and $T_{N2}$ with $x$ is also reflected in specific heat measurements (Fig. 1c). The lattice contribution in all cases can be described by $C_{p} \propto T^{2}$, an approximation of the 2D Debye law characteristic of the parent compound \cite{banerjee_proximate_2016} and other van der Waals bonded materials \cite{krumhansl_lattice_1953}. The 2D character of the materials is thus not strongly affected by Ir substitution. Weak low temperature features with a magnetic origin (evidenced by magnetic field dependence) are observed at larger $x$ of 0.2 and 0.3 [see SI, Fig. S9], however the absence of a $\lambda$ - like anomaly indicates no long-range ordering transition above 2~K. 

\begin{figure}[t]
\centering
\includegraphics[scale=0.43]{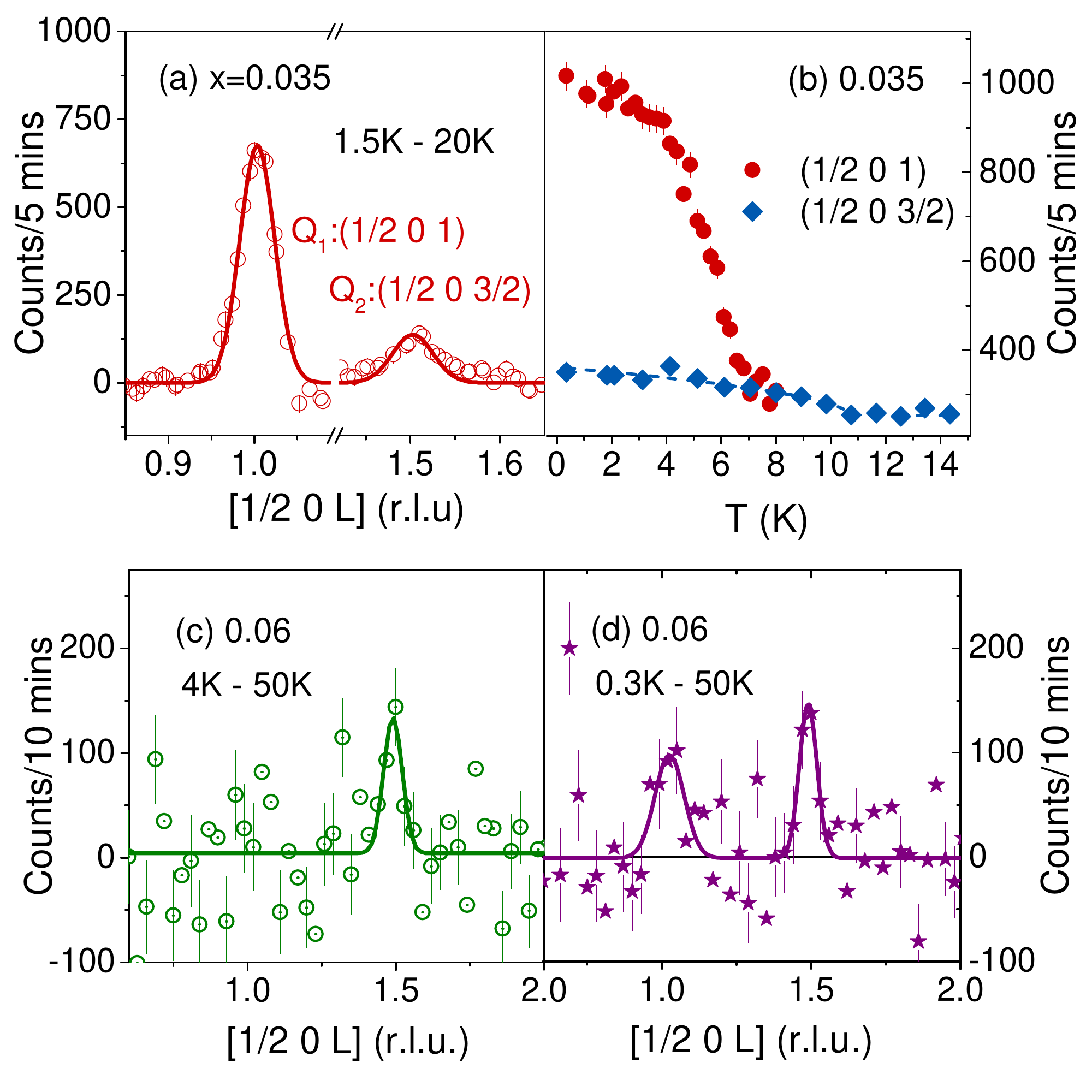}
\caption{(a) Scan at 1.5 K collected at HB-1A along [1/2 0 L] through characteristic magnetic reflections \textbf{Q}$_{1}= $(1/2 0 1) and \textbf{Q}$_{2}= $(1/2 0 3/2) for a 50~mg single crystal with $x = 0.035$. 20 K ($T > T_{N}$) data are subtracted as a background. (b) Temperature scans of the scattering intensity at  \textbf{Q}$_{1}$ and \textbf{Q}$_{2}$. Scans along [1/2 0 L] at (c) 4~K and (d) 0.3~K in a 20~mg single crystal with $x = 0.06$. 50 K data are subtracted as a background.} 
\end{figure}

\begin{figure}[t]
\centering
\includegraphics[width=1.0\linewidth]{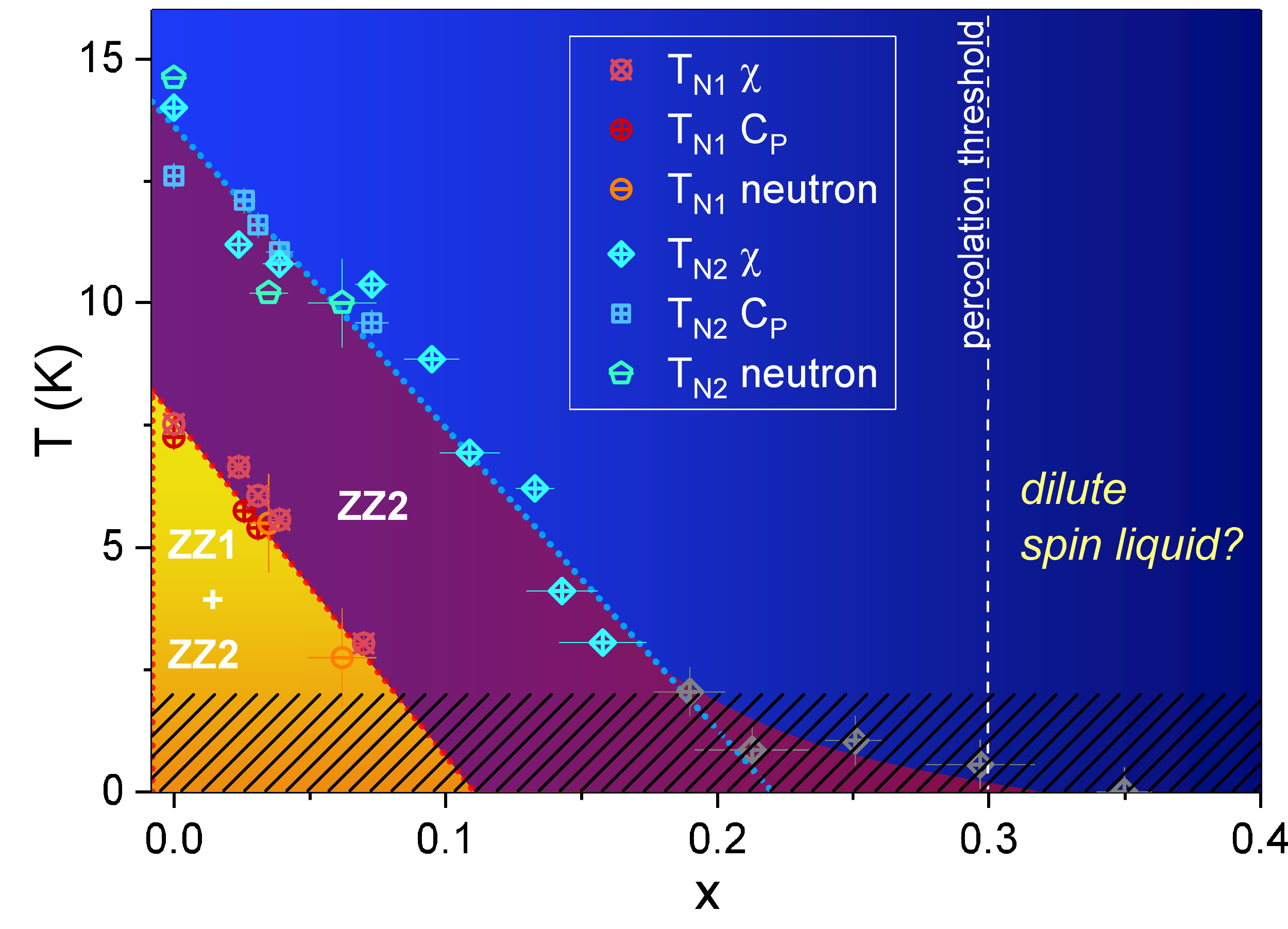}
\caption{Phase diagram of Ru$_{1-x}$Ir$_{x}$Cl$_{3}$. Transition temperatures $T_{N1}$ (red symbols) and $T_{N2}$ (blue symbols) determined from magnetic susceptibility, heat capacity, and neutron diffraction mark the boundaries of the ABC-stacked (ZZ1) and ABAB-stacked (ZZ2) zigzag phases in multiphase single crystals. Linear fitting to $T_{N}(x)$ (dotted lines) gives a critical dilution level for $T_{N1} \rightarrow 0$ of $x \sim 0.11$. The cusp at $T_{N2}$ is driven below the experimental base temperature of 2~K (indicated by hashmarks) by $x \simeq 0.2 $; at larger $x$ $T_{N2}$ is estimated by extrapolation to $d\chi/dT =0$ (grey symbols).} 
\end{figure}

\begin{figure}[t]
\centering
\includegraphics[width=1.01\linewidth]{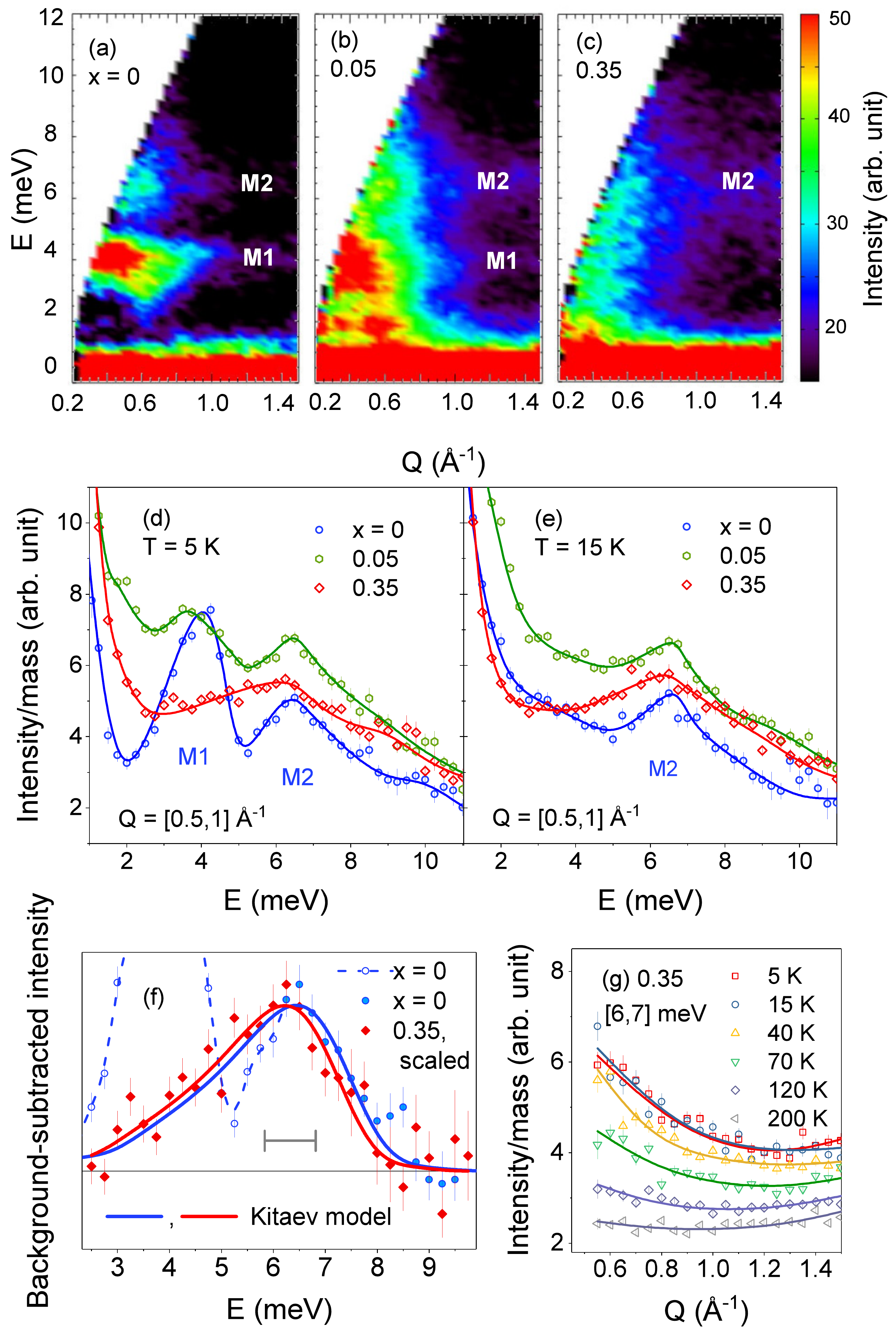}
\caption{ Powder inelastic scattering of Ru$_{1-x}$Ir$_{x}$Cl$_{3}$ measured at SEQUOIA with Ei = 25 meV. The background from Al sample cans is subtracted from all data, and a correction for the increasing absorption cross section with Ir content is applied to allow direct comparison of intensites. (a)-(c) 2D spectra at 5 K ($x$ values indicated). Constant-Q cuts integrated over Q = [0.5,1]\AA $^{-1}$ are shown at (d) 5 K and (e) 15 K. Solid lines are a guide to the eye. (f) Constant-Q cuts at 5 K for $x=0$ and $x=0.35$ after subtraction of a linear background. The $x = 0.35$ data have been scaled by a factor of 1.6 to match the height of the $x=0$ data. Solid blue and red lines represent the  calculated scattering of the upper mode of the Kitaev model convoluted with experimental resolution (grey bar) for $x =0$ and $x = 0.35$, respectively. At low energies the scattering in the magnetically ordered parent compound departs from the model (open symbols, dashed line). (g) Constant-energy cuts at the M2 peak position of [6,7] meV for $x=0.35$ normalized by $n(\omega) + 1$ to account for the temperature dependence of phonon contamination at low Q. Solid lines are a guide to the eye.} 
\end{figure}

The magnetic ground state of lightly-substituted crystals with $x \leq 0.06$ was characterized by neutron diffraction using the HB-1A and HB-3A beamlines at HFIR, ORNL. Similar to the parent compound with stacking faults, Ir substituted crystals show transitions at $T_{N1}$ with (trigonal notation) \textbf{Q}$_{1}$ = (1/2 0 1) and  $T_{N2}$ with \textbf{Q}$_{2} = \left( 1/2\ 0\ 3/2\right)$ \cite{banerjee_neutron_2017,banerjee_proximate_2016}, corresponding, respectively to ABC and ABAB magnetic layer stacking. For $x = 0.035$ (Fig. 2a,b) the relative peak intensities $I$(\textbf{Q}$_{1}$)/$I$(\textbf{Q}$_{2}$) $\simeq 6$ at 1.5~K show that the majority of the sample adopts the ABC-type phase.  The observed $T_{N1}$ of $\simeq$ 5.5~K is consistent with $\chi$ and $C_{p}$ for similarly doped samples.  Measurements using the four-circle diffractometer at HB-3A were refined to yield an ordered moment of 0.32(2)~$\mu_{B}$/Ru in the ABC zigzag phase, corresponding to two-thirds of the ordered moment at $x=0$  \cite{cao_low-temperature_2016}. An extensive survey of reciprocal space did not reveal additional peaks associated with alternative magnetically ordered states in the H-K-$\Gamma$ phase diagram \cite{rau_generic_2014}.

Increasing the Ir content to $x = 0.06$ (Fig. 2c,d) strongly affects $T_{N}$ as well as the relative intensities at the two magnetic wavevectors. The (1/2 0 1) peak is absent at 4~K (Fig. 2c) and acquires significant intensity only below 1.6 K. The relative amount of the ABC phase is greatly reduced, leading to $I$(\textbf{Q}$_{1}$)/$I$(\textbf{Q}$_{2}$) $\simeq 0.7$ at 0.3 K. No additional peaks characteristic of other candidate magnetic structures were observed in the (H 0 L) scattering plane. There is thus no evidence for the emergence of a new magnetically ordered phase as the ABC-type zigzag order is suppressed in the dilute system. 

A phase diagram of the Ru$_{1-x}$Ir$_{x}$Cl$_{3}$ system is summarized in Fig. 3. Linear fits to the normalized critical temperatures give initial suppression rates \cite{cheong_magnetic_1991} $-d[T_{Ni}/T_{Ni}(0)]/dx = 8.1(7)$ for $T_{N1}$ and 4.4(3) for $T_{N2}$. The more sensitive response to magnetic site dilution in the ABC phase may be related to a higher degree of frustration in magnetic interactions for 3-layer magnetic stacking than for 2-layer stacking, as evidenced by the different $T_{N}$ values for the two phases in the parent compound. Extrapolating linearly from small $x$, the suppression of $T_{N1}$ and $T_{N2}$ to $T=0$ occurs at $x \simeq 0.11$ and $x \simeq 0.22$, respectively. Above $x \simeq 0.2$, $T_{N2}$ falls below the base temperature of 2 K for susceptibility measurements. However, extrapolation of $d\chi/dT$ to zero indicates that $T_{N2}$ may not vanish until the percolation threshold is crossed (grey symbols, Fig. 3). 

The nature of the magnetic ground state in Ru$_{1-x}$Ir$_{x}$Cl$_{3}$ can be further revealed via examination of the excitation spectrum using inelastic neutron scattering (INS). The existing Ir-substituted single crystals are too small for INS experiments and have coexisting magnetic ground states.  On the other hand, powders exhibit only the ABAB type ordering \cite{banerjee_proximate_2016}.  The 14~K ($T_{N2}$-type) transition in polycrystalline RuCl$_{3}$ becomes suppressed by $x \simeq 0.3$, similar to the trend in $T_{N2}$ for the single crystals (SI, Fig. S3). The SEQUOIA time-of-flight spectrometer at the Spallation Neutron Source (SNS) was used to measure the response of powders with $x=0$, 0.05, and 0.35 (Fig. 4). The INS spectrum of RuCl$_{3}$ ($x=0$) shows two magnetic features \cite{banerjee_proximate_2016} (Fig. 4a). The lower feature M1 arises from zigzag ordered state spin waves and vanishes above $T_{N2}$. In contrast, the temperature-dependence and energy width of the upper feature M2 $\simeq$6 meV are incompatible with spin-wave theory, but resemble the calculated spectrum for fractionalized excitations of the pure Kitaev QSL model \cite{banerjee_proximate_2016,banerjee_neutron_2017,knolle_dynamics_2014}.

Ir substitution affects M1 most strongly as seen in Fig. 4(a-c). The energy dependence of the scattering intensity normalized by sample mass and integrated over the Q range [0.5,1]~\AA $^{-1}$ is shown at $T = 5$ K (Fig. 4d) and $T = 15$ K (Fig. 4e).  At 5 K the modes M1 and M2 are clearly visible for $x=0$, and M1 is gapped.  For $x=0.05$, with $T_{N2} \simeq $ 10 K, both modes are also observed, although M1 is broadened and renormalized downwards, and no M1 gap is visible as significantly more scattering is observed at low frequencies.  In contrast, the $x = 0.35$ sample, which is magnetically disordered, shows only one broad feature peaked near M2. At 15 K, above $T_{N2}$ for both $x=0$ and $x=0.05$, the M1 peak disappears as expected since it arises from spin waves associated with zigzag order.  On the other hand, for all three concentrations the M2 feature remains robust, with very little temperature dependence. 

Figure 4(f) shows the constant-Q cuts for $x=0$ and $x=0.35$ at 5 K after the subtraction of a linear background.  The width of the M2 scattering feature is very broad compared with instrumental resolution.  It was shown previously that the scattering on the high-energy side of the M2 mode for $x=0$ \cite{banerjee_proximate_2016} bears a strong resemblance to that calculated for the isotropic antiferromagnetic Kitaev model at Q = 0 \cite{knolle_dynamics_2014} (blue line, Fig. 4f). The red line in Fig. 4(f) is a fit of the $x=0.35$ data to this model over the full energy range of the M2 feature, showing very good agreement with a value of the Kitaev energy scale $K =5.3(2)$ meV.  

The Q dependence of the scattering in the vicinity of the M2 mode in the $x=0.35$ system is illustrated in Fig. 4(g), with the data normalized by $(1-e^{-E/k_{B}T})^{-1}$.  The signal exhibits little change up to moderate temperatures, consistent with expectations for a QSL with the energy scale set by $K$ \cite{nasu_thermal_2015}. However, for higher temperatures the scattering shows a stronger temperature dependence than the pure material \cite{banerjee_proximate_2016, banerjee_neutron_2017}, suggesting that the excitations in the dilute system are more fragile.

The energy, temperature, and wave-vector dependence of the spectrum at $x=0.35$ are all consistent with expectations for fractionalized excitations, suggesting that the $T=0$ ground state is effectively a QSL.  In the pure Kitaev model spin-spin correlations do not extend beyond nearest neighbors, and thus are relatively insensitive to non-magnetic dilution. Therefore it is reasonable that these correlations and their associated excitations exist beyond the percolation threshold for long-range order.  Excitations in Kitaev systems with mobile holes have been predicted to form bound states \cite{halasz_doping_2014, halasz_coherent_2016}, which one might expect to manifest as sharp peaks in the magnetic response function.  No such peaks are observed in the present experiment, however this is not surprising as the nonmagnetic Ir$^{3+}$ impurities here are completely static and do not introduce any change in charge neutrality to the system.

In summary, incorporating Ir$^{3+}$ into Ru$_{1-x}$Ir$_{x}$Cl$_{3}$ decreases the N\'eel temperatures of the ordered magnetic phases, while leaving intact the broad upper excitation mode associated with fractionalized excitations in the $x=0.05$ and $x=0.35$ compositions investigated by INS.  Above the percolation threshold where long-range order is absent, the fractionalized excitations dominate the spectrum and the low-temperature region may effectively be a dilute QSL. The robust nature of such QSL physics in RuCl$_{3}$ with respect to chemical substitution strongly motivates further investigation of dopants, in particular those that would introduce mobile charge carriers, an avenue which is predicted to bring about exotic superconductivity \cite{hyart_competition_2012, you_doping_2012, mei_possible_2012, okamoto_doped_2013, okamoto_global_2013, trousselet_hole_2014, kimme_symmetry-protected_2015}.

\textbf{Acknowledgments}

P.L.K and D.M. were supported by the Gordon and Betty Moore Foundations EPiQS Initiative Grant GBMF4416. J.-Q.Y. and C.A.B. acknowledge support from the U.S. DOE, Office of Science, Basic Energy Sciences, Materials Sciences and Engineering Division. The work at ORNL HFIR and SNS was sponsored by the Scientific User Facilities Division, Office of Basic Energy Sciences, U.S. DOE.  We thank I. McKenzie for support at TRIUMF, B. Chakoumakos and F. Ye for assistance with XRD, and B. Sales, M. McGuire, and A. May for helpful discussions. 


%

\clearpage

\renewcommand{\thefigure}{S\arabic{figure}}

\setcounter{figure}{0}

\renewcommand{\thetable}{S\arabic{table}}

\setcounter{table}{0}

\begin{figure*}[t]
\centering
\includegraphics[scale=0.425]{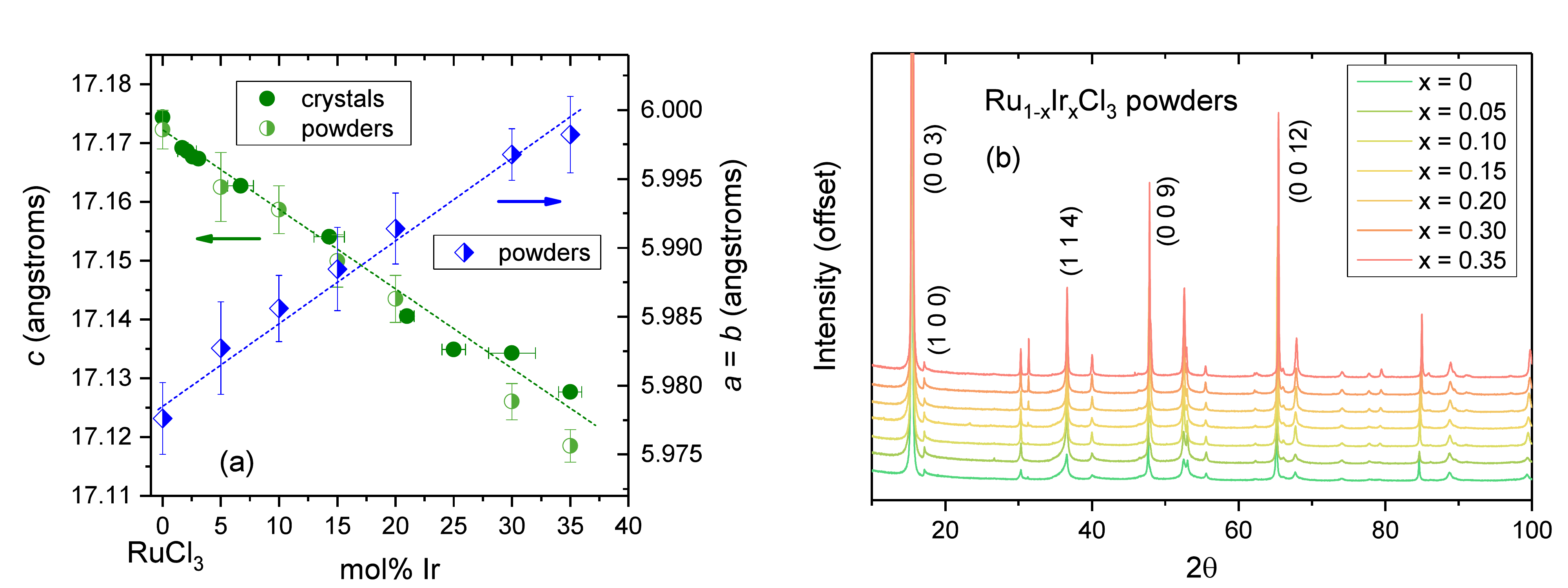}
\caption{Fig. S1 (a) Room temperature lattice parameters of Ru$_{1-x}$Ir$_{x}$Cl$_{3}$ single crystals (filled symbols) and powders (half-filled symbols) indexed in the trigonal P3$_{1}$12 setting. Lines are a guide to the eye. (b) Offset X-ray diffraction patterns of Ru$_{1-x}$Ir$_{x}$Cl$_{3}$ powders. Average Ir content and distribution is determined by EDS measurements in single crystals; Ir content in the powdered materials is taken as the nominal value. } 
\end{figure*}

\subsection{I. Experimental details}

Single crystals of Ru$_{1-x}$Ir$_{x}$Cl$_{3}$ were grown by a transport method at high temperatures (900-1100$^{\circ}$C) from commercial powders of RuCl$_{3}$ (Furuya Metals) and IrCl$_{3}$ (American Elements) in the presence of NaCl. Polycrystalline Ru$_{1-x}$Ir$_{x}$Cl$_{3}$ was synthesized from direct reaction of purified RuCl$_{3}$ powder pelletized with IrCl$_{3}$ and sealed in an evacuated quartz tube with the low-decomposing chloride TeCl$_{4}$ as a source of Cl$_{2}$ atmosphere to minimize oxide formation.

Single crystal neutron diffraction was carried out at the HB-1A triple axis spectrometer ([H 0 L] scattering plane) and HB-3A four-circle diffractometer at the High Flux Isotope Reactor at ORNL with incident neutron wavelengths of 2.36 and 1.546 \AA, respectively. Inelastic neutron scattering was measured at the Spallation Neutron Source using the SEQUOIA chopper sepctrometer for powders of Ru$_{1-x}$Ir$_{x}$Cl$_{3}$ with $x=0,0.05$ and 0.35 and total masses of 5.3, 4.97, and 4.92g, respectively. The $x=0$ sample was placed in a flat-plate Al sample holder, while the Ir-substituted compositions were measured in annular Al sample cans due to the large absorption cross section of Ir. Data were surveyed at several incident neutron energies between 6 and 100 meV, with focus on 8 meV and 25 meV to cover the features of interest with adequate resolution. Muon spin rotation experiments on powder samples were performed at the TRIUMF M20 beamline in the LAMPF $\mu$SR spectrometer.

\subsection{II. Structure and composition}

Single crystal X-ray diffraction (Rigaku, Mo K$\alpha$ radiation) was performed for the compositions $x = 0.05$ and $x = 0.15$ on small crystals with $\sim 100 \mu$m in-plane dimension. The structure was refined using SHELX to the monoclinic C2$/m$ space group, consistent with similarly-sized RuCl$_{3}$ single crystals and the reported space group of $\alpha$-IrCl$_{3}$ [1], with Ir substituting for Ru on the $4g$ site.  Individual planes of the larger ($\sim$mm) sized plate-like crystals used in SQUID and heat capacity measurements were checked by X-ray diffraction on a Panalytical Xpert Pro powder diffractometer. Indexed in the trigonal P3$_{1}$12 setting, the out-of-plane lattice parameter of these crystals as determined from the (0 0 L) peaks decreases linearly with increasing Ir substitution (Fig. S1a). In powders, XRD patterns (Fig. S1b) show strong (0 0 L) texturing in a fraction of the samples, and Warren lineshapes are characteristic of the quasi-2D nature of the system. The c-axis lattice parameters determined from powder XRD are in good agreement with those of the single crystal samples while the in-plane lattice parameter grows systematically with incorporation of Ir (Fig. S1a). The unit cell volume also increases approximately linearly with increasing $x$ from 531.4(3) \AA$^{3}$ to 533.4(2)\AA$^{3}$.

\begin{figure}
\centering
\includegraphics[scale=0.8]{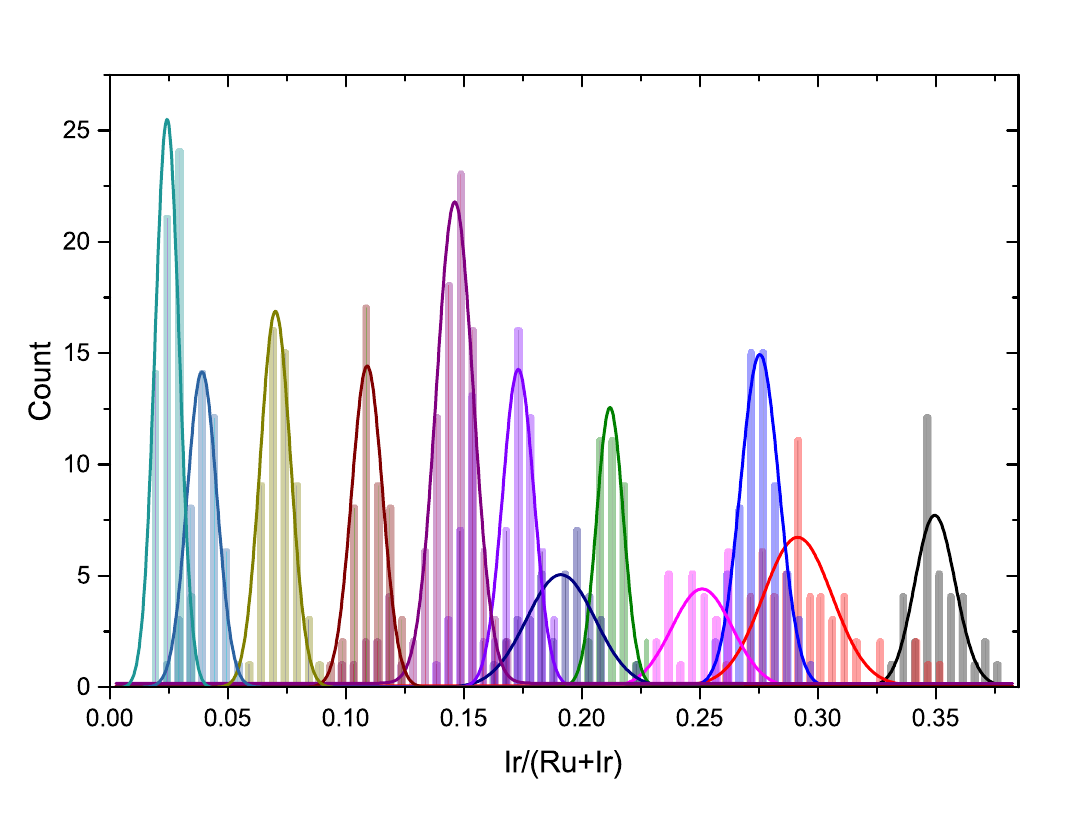}
\caption{Histograms representing the substitution level Ir/(Ru+Ir) over a number of sites in single crystals of Ru$_{1-x}$Ir$_{x}$Cl$_{3}$ as determined by EDS measurements. Solid lines represent a Gaussian fit to the distribution in each sample. } 
\end{figure}

The elemental compositions of the Ru$_{1-x}$Ir$_{x}$Cl$_{3}$ single crystals in this study were checked individually by energy dispersive X-ray spectroscopy (EDS). Variation of Ru$/$Ir ratio over $\sim$mm length scales was noted in samples large enough for bulk measurements; the distribution of $x$ was characterized in each sample by measuring a large number of sites spanning the crystal with a spot size of $20-30 \mu$m.  Representative histograms for twelve crystals are shown in Fig. S2. Standard deviation from the mean value of $x$ ranged from 0.3 mol \% Ir to 2 mol\% Ir, but remained less than 1\% in nearly all of $>$20 crystals tested. For the powder samples, $x$ indicates the nominal composition.

\subsection{III. Powder susceptibility}
The magnetic susceptibility of Ru$_{1-x}$Ir$_{x}$Cl$_{3}$ powders measured in a field of $B = 1$~T is shown in Fig. S3. As seen for  $T_{N2}$ in the single crystals, increasing $x$ continuously depresses the susceptibility cusp marking the onset of the zigzag phase, and $T_{N2}$ initially varies pseudo-linearly with $x$ with a slower fall-off as the percolation threshold is approached. However, we note that the linear extrapolation of the critical line from small $x$ gives a critical concentration of $x_{c} \sim 0.16$, between the critical values of 0.11 and 0.22 for $T_{N1}$ and $T_{N2}$, respectively, in the single crystals. The initial suppresstion rate of $-d[T_{Ni}/T_{Ni}(0)]/dx =7.6$ is also intermediate between the values for the single crystal transitions, although it is close to the $T_{N2}$ value of 8.2. The differentiation between $T_{N2}$ in the powders and single crystals with dilution highlights the sensitivity of the three dimensional ordered state to layer stacking, which is ABAB ordered in the crystals while the stacking sequences in the polycrystalline material are random.   

As observed in the single crystals, the concavity of the field-dependent magnetization curves also changes from concave up to concave down with increasing Ir content (Fig. S3b). We note that the magnitude of both temperature- and field-dependent magnetization at low temperatures evolves non-monotonically with $x$, showing an increase in moment over an intermediate range, roughly $0.1 < x < 0.2$, while decreasing elsewhere. This observation is consistent with the trends shown for the single-crystal magnetization in Fig. 1 of the manuscript and Supplementary Fig. S8, indicating that the origin of this behavior is intrinsic and relatively insensitve to structural details.

\begin{figure}
\centering
\includegraphics[scale=0.38]{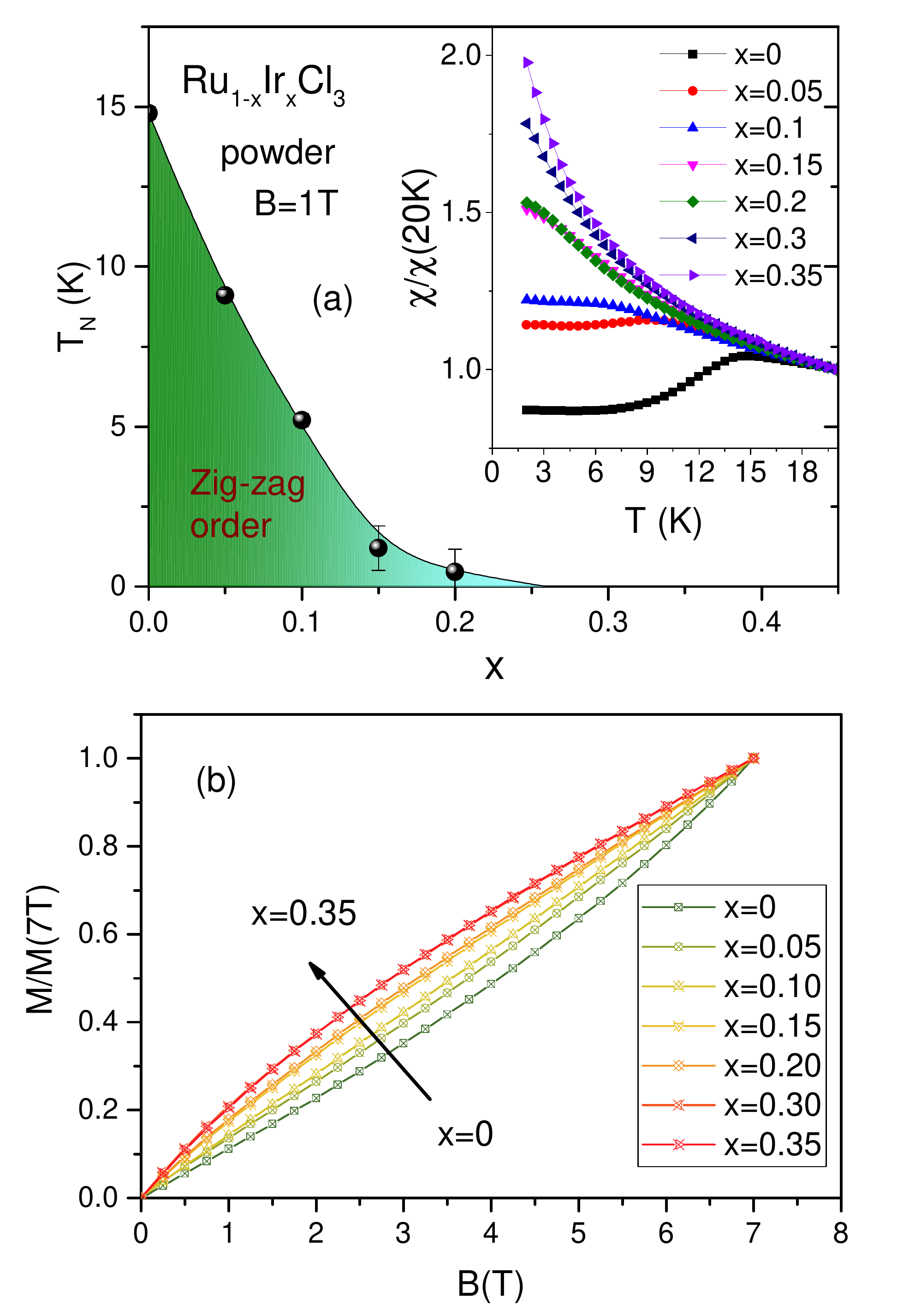}
\caption{(a) Magnetic transition temperatures $T_{N2}$ vs $x$ for polycrystalline Ru$_{1-x}$Ir$_{x}$Cl$_{3}$ determined from the cusp in temperature-dependent susceptibility. Critical points below the experimental base temperature of 2~K are extrapolated (See section VI). Inset shows $\chi (T)$ curves measured in $B = 1$~T for various $x$, normalized to their 20~K value. (b) Field-dependent magnetization curves at 2~K for various compositions, normalized to their $B = 7$~T value. } 
\end{figure}

\subsection{IV. Muon spin spectroscopy}

$\mu$SR measurements were performed on Ru$_{1-x}$Ir$_{x}$Cl$_{3}$ powders to characterize the weakly magnetic state of the dilute system.  The relaxation of the muon asymmetry $a(t)$ reflects the amplitudes and fluctuations of local magnetic fields [2]. Relaxation of the muon asymmetry due to nuclear moments in a polycrystalline sample is given by the well- known Kubo-Toyabe function,
\begin{equation}
a_{KT} = a_{0} \left[ 1/3 + 2/3 \left(1-\Delta^{2} t^{2} \right) exp\left(-\Delta^{2} t^{2}\right) \right]
\end{equation}

\begin{figure*}
\centering
\includegraphics[scale=0.4]{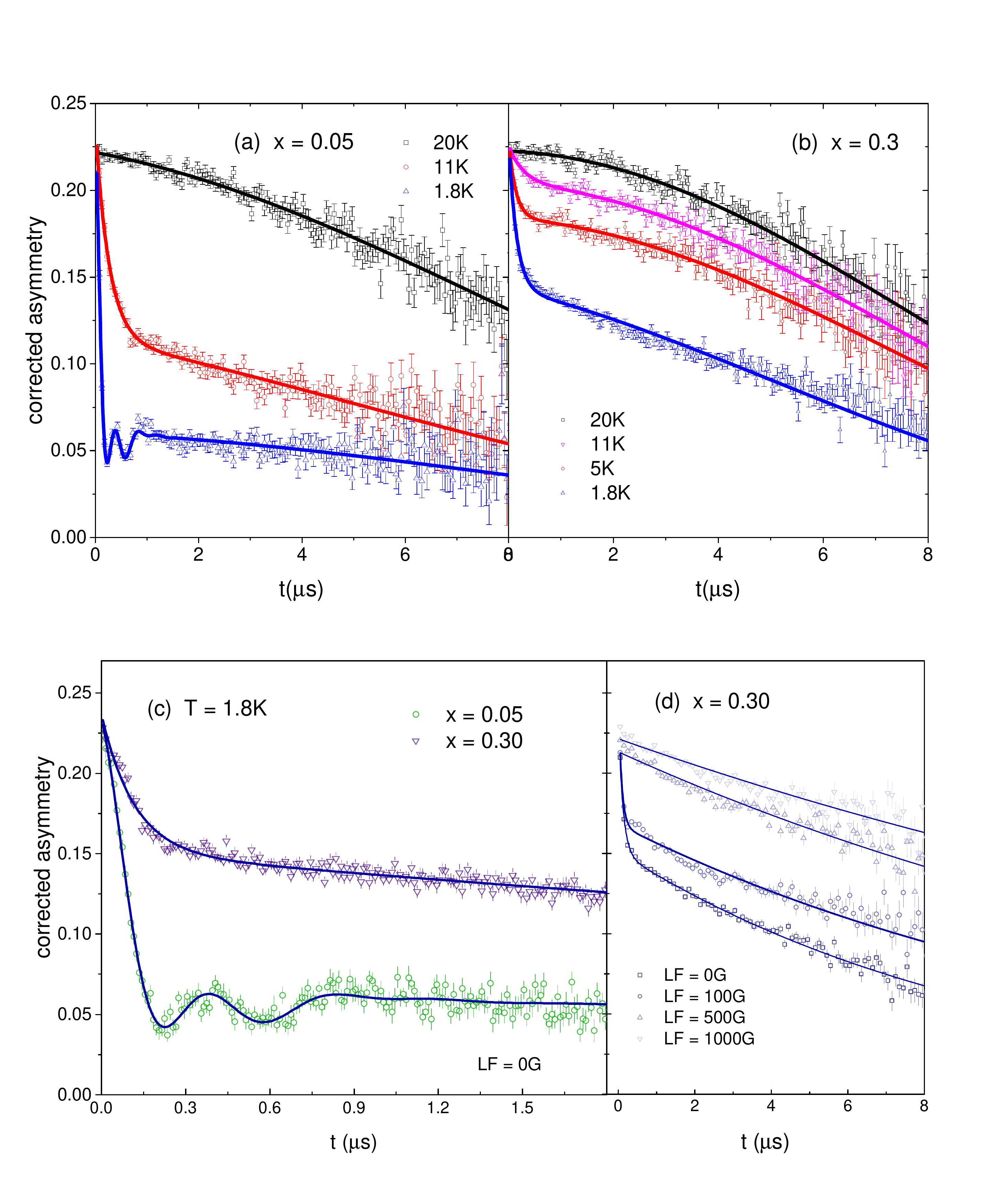}
\caption{ Zero-field $\mu$SR spectra at various temperatures in Ru$_{1-x}$Ir$_{x}$Cl$_{3}$ powders with (a) $x = 0.05$ and (b) $x = 0.30$. Solid lines are fits to relaxation models described in the text. (c) Comparison of 1.8 K spectra for $x = 0.05$ and  $0.30$ at short times. (d) Longitudinal field (LF) curves for $x = 0.30$ at 1.8~K show a decoupling of the fast component for LF $\geq 500$~G. Persisting exponential relaxation in the LF spectra indicate dynamic spin fluctuations.} 
\end{figure*}

\begin{table*} [t]
\begin{center}
 \begin{tabular}{||c c c c c c c c c||} 
 \hline
$x = 0.05$ & $A_{0}$ & $\lambda_{0}$ ($\mu$s$^{-1}$) & $A_{1}$ & $\lambda_{1}$ ($\mu$s$^{-1}$) & $\omega_{1}$ (MHz) & $A_{2}$ & $\lambda_{2}$ ($\mu$s$^{-1}$) & $\omega_{2}$ (MHz)\\ [0.5ex] 
 \hline\hline
 $T = 11$ K & 0.543(5)&	0.089(3)	&0.54(1)&	4.1(2)&	-	&-	&-	&-\\ 
 \hline
 $T = 1.8$ K & 0.284(4) & 0.057(5) & 0.37(4) & 4.4(4) & 2.40(4) & 0.41(4) & 3.8(4) & 0.92(3)\\ [1ex] 
 \hline
\end{tabular}
\caption{Fitted parameters of the $\mu$SR spectra of polycrystalline Ru$_{1-x}$Ir$_{x}$Cl$_{3}$ with $x = 0.05$, see Eqn. (2).} 
\end{center}
\end{table*}

where $\Delta$ is the width of the local distribution of static fields. At 20 K in the Ru$_{1-x}$Ir$_{x}$Cl$_{3}$ powders (Fig. S4a,b), weak Gaussian relaxation can be described by Eqn. 1, with $\Delta  = 0.093(1)$ and 0.095(9) $\mu$s$^{-2}$ for $ x = 0.05$ and $x = 0.3$, respectively. In the sample with $x = 0.05$, a fast-relaxing component emerges near $T_{N2}$  at 11 K and well-defined oscillations develop by 1.8 K (Fig. S4a). Two distinct frequencies $\omega_{1}/2\pi\sim 2.5$~MHz and $\omega_{2}/2\pi\sim 1$~MHz are observed, in agreement with a recent report in $\alpha$-RuCl$_{3}$ powders that found two inequivalent muon sites in the pseudo-bilayer zigzag phase [3]. The asymmetry is generally modeled as
\begin{equation}
a(t)=a_{0} [A_{0} exp(-\lambda_{0} t)+\Sigma_{(i=1,2)} \left[ A_i exp(-\lambda_{i} t)cos(2\pi\omega_{i} t) \right]                    
\end{equation}
where $A_{i}$ represent the fast-relaxing/oscillating components and $A_{0}$ is a slowly relaxing exponential tail. The fitted parameters for the $x = 0.05$ sample are given in Table S1. At 1.8 K, the total oscillating volume fraction $A_{1}+A_{2} = 0.75(6)$ is close to the expected value of 2/3 for a fully ordered long-range magnetic system. 

In contrast, at $x = 0.30$ no coherent muon precession frequencies that would indicate long-range order are resolved at any temperature down to 1.8 K (Fig. S4b). A two-component Gaussian plus rapidly-decaying exponential spectrum is observed at both 11 K and 5 K, and is modeled as 
\begin{equation}
a(t)=a_{KT}[A_{0}+A_{1} exp(-\lambda_{1} t)].
\end{equation}
At 1.8~K the spectrum is comprised of two exponential terms $A_{i} e^{-\lambda_{i} t}$. An exponential functional form of the muon relaxation is consistent with the fast-fluctuation limit of distributed internal magnetic fields [2]. The two-component exponential spectrum at base temperature is modeled by 
\begin{equation}
a(t)=a_{0}[A_{0} exp(-\lambda_{0} t)+A_{1} exp(-\lambda_{1} t)].
\end{equation}
The fitted parameters are given in Table S2. 

Longitudinal field experiments, in which spin depolarization due to static internal fields is suppressed, (Fig. S4d) were used to distinguish between static and dynamic processes in the $x = 0.30$ sample. A static origin can be assigned to the early rapid depolarization in the zero-field spectrum ($\lambda_{1}\sim $ 9~MHz), which is decoupled in a longitudinal field LF $\geq 500$~G (Fig. S4d). However, the persistence of the slow exponential decay term $\lambda_{2}\sim $ 0.1~MHz in the LF spectra is a clear demonstration of dynamic magnetism in the fast-fluctuation limit in the remaining volume of the sample. It is unknown whether or not these two components are also present in the single crystals.  

\begin{table}[t]
\begin{center}
 \begin{tabular}{||c c c c c c||} 
 \hline
$x = 0.30$ & $\Delta$ & $A_{1}$ & $\lambda_{1}$ ($\mu$s$^{-1}$) & $A_{2}$ & $\lambda_{2}$ ($\mu$s$^{-1}$)\\ [0.5ex] 
 \hline\hline
 $T = 20$ K & 0.095(9) & - & -& - & -\\ 
 \hline
 $T = 11$ K & 0.096(1) & 0.109(7) & 2.8(4)& 0.914(3) & -\\ 
 \hline
 $T = 5$ K & 0.097(1) & 0.209(1) & 6.1(6)& 0.823(2) & -\\ 
 \hline
 $T = 1.8$ K & - & 0.39(1) & 8.6(4) &0.687(3)  & 0.1007(2)\\ [1ex] 
 \hline
\end{tabular}
\caption{Fitted parameters of the $\mu$SR spectra of polycrystalline Ru$_{1-x}$Ir$_{x}$Cl$_{3}$ with $x = 0.3$, see Eqn. (3)-(4).} 
\end{center}
\end{table}

\subsection{V. Excitation gap}

To examine the low-energy inelastic scattering in detail, INS spectra were acquired using an incident neutron energy of 8 meV, which provides a resolution of 0.12 meV at the elastic line. Fig. S5(a) and (b) show the 5 K spectra for $x=0$ and $x=0.05$, respectively. A sharp spurious feature at $\sim 0.7$ meV arises in both spectra due to multiple scattering involving the sample environment. In the parent compound, the spin-wave mode softens at the M-point of the Brillouin zone, Q = 0.6\AA~$^{-1}$, with a concave low-Q edge consistent with the spin waves of the zigzag state. The absence of structured scattering below 1.7 meV clearly points to a gapped spin-wave spectrum. In the lightly-substituted $x=0.05$ material, the spin-wave mode softens strongly at the M-point, with states filling in the formerly gapped region. Fig. S5(c,d) shows constant-Q cuts through the M-point Q = [0.5, 0.7] \AA$^{-1}$ and away from the mode Q = [1.1, 1.4] \AA$^{-1}$. Linear extrapolation of the leading edge of the mode indicates that it reaches the background level very close to $E = 0$, indicating a gapless or nearly gapless spectrum in the magnetically ordered region of the phase diagram with $x \gtrsim 0.05$. 

\begin{figure*}
\centering
\includegraphics[scale=0.5]{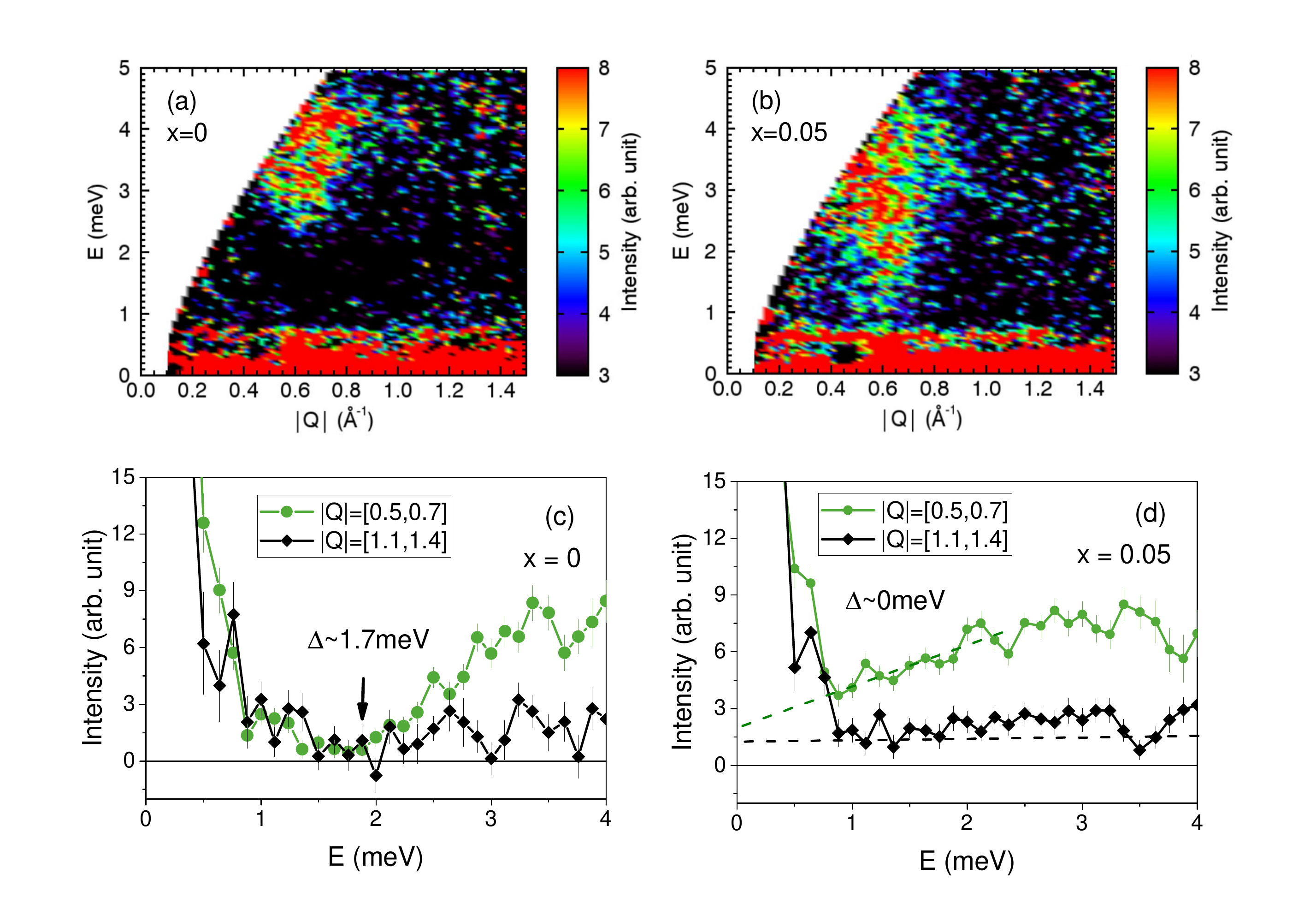}
\caption{Inelastic neutron scattering spectra for (a) $x=0$ and (b) $x=0.05$ acquired at 5~K using an incident neutron energy of Ei = 8 meV. Constant-Q cuts of the data in (a) and (b) through the spin-wave mode ([0.5, 0.7] \AA $^{-1}$) and a background position ([1.1, 1.4] \AA $^{-1}$) are shown in panels (c) and (d), respectively. Green dashed line is a linear extrapolation of the scattering from the spin wave mode. Black dashed line in (d) indicates the background level.} 
\end{figure*}

\begin{figure*}
\centering
\includegraphics[width=0.9\linewidth]{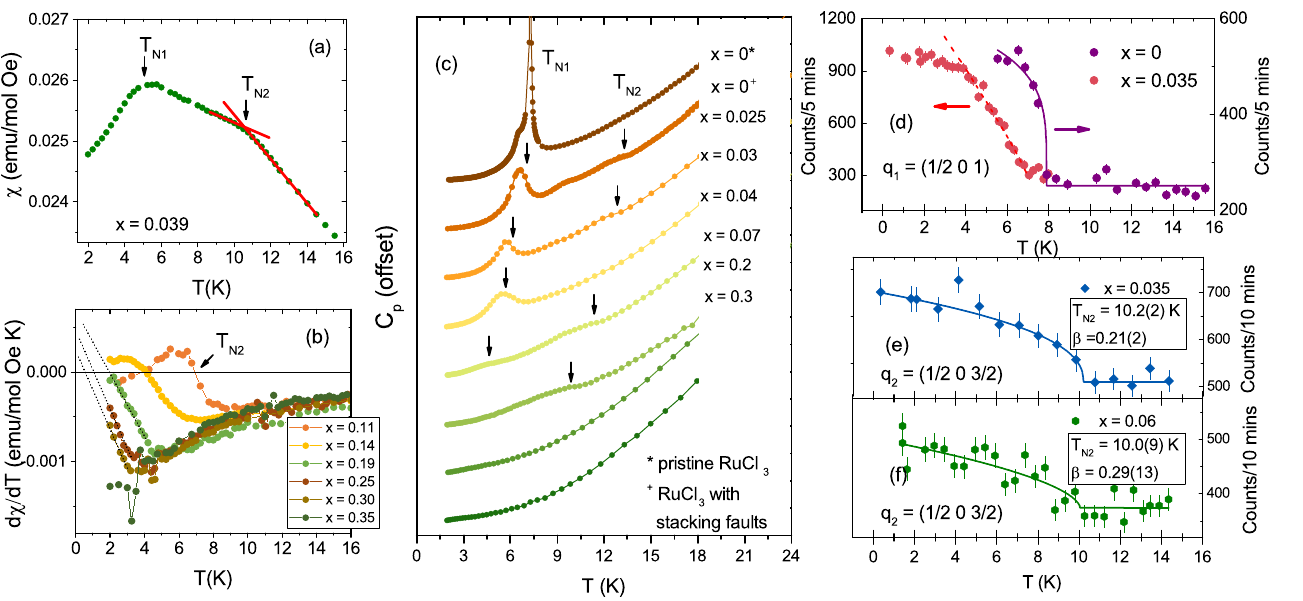}
\caption{(a),(b) Determination of magnetic transition temperatures from magnetic susceptibility curves. Solid lines in (b) represent linear extrapolation of $d\chi / dT$ to estimate $T_{N2}$ below the experimental base temperature of 2 K. (c) Offset heat capacity curves for various Ir substitution levels. Arrows indicate $T_{N1}$ and $T_{N2}$. (d) Intensity vs. temperature for the (1/2 0 1) magnetic Bragg peak appearing at $T_{N1}$ in $x = 0$ and $x = 0.035(7)$ single crystals. The substituted crystal shows a smearing of the transition onset compared to the parent compound, and intensity is not well described a power law. Dashed line is a guide to the eye. Solid line is a fit to $I =I_{0} + A(1-\frac{T}{T_{N}} )^{2\beta}$ for $x = 0$. (e),(f) Intensity vs. temperature for the (1/2 0 3/2) magnetic Bragg peak appearing at $T_{N2}$ for (e) $x = 0.035(7)$ and (f) $x = 0.062(13)$ single crystals. Solid lines are power law fits. } 
\end{figure*}

\subsection{VI. Single crystal magnetic characterization}

\subsubsection{Curie Weiss behavior}
Large effective paramagnetic moments of the $S = 1/2$ Ru ions in $\alpha$-RuCl$_{3}$ have been previously noted and are attributed to a mixed spin-orbital character [4]. For Ru$_{1-x}$Ir$_{x}$Cl$_{3}$, the effective paramagnetic moment per formula unit determined from Curie-Weiss fittings of the high-temperature susceptibility data presented in Fig. 1(a) decreased from 2.4 $\mu_{B}$ at $x=0$ to 1.9 $\mu_{B}$ at $x=0.30$, consistent with S = 0 Ir ions within systematic uncertainties. Normalized to Ru content, the moment per Ru ion in Ru$_{1-x}$Ir$_{x}$Cl$_{3}$ retains large values in the range of 2.3 - 2.7 $\mu_{B}$/Ru$^{3+}$; a significant trend with $x$ cannot be established from the current experiment, however some small systematic dependence also cannot be ruled out. Curie-Weiss intercepts $T_{CW}$ are positive in all samples and follow an overall decreasing trend with $x$, although the values appear to level off at $T_{CW} \sim 20~$K for $x > 0.1$ after initially dropping more rapidly. 

\subsubsection{Magnetic transition temperatures}
Magnetic transition temperatures were estimated either from the cusp ($\frac{d\chi}{dT}=0$) in the temperature-dependent susceptibility curves or by linear extrapolation of data above and below the turnover in $\chi$ for cases in which no well-defined cusp exists, e.g. $T_{N2}$ in 2-phase samples. Critical temperature determination from susceptibility is illustrated in Fig. S6(a,b). These transition temperatures are in good agreement with heat capacity and neutron diffraction data collected for select compositions (Fig. S6c-f). 

\subsubsection{Hysteresis}

To check for the presence of spin-glass-like freezing and/or thermal hysteresis at low temperatures, susceptibility was measured in a magnetic field of 1 T under zero-field-cooled (ZFC), field-cooled cooling (FCC) and field-cooled warming (FCW) protocols for several compositions spanning the range investigated. No splitting was observed between the ZFC/FC curves in Fig. S7(a) or in FCC/FCW curves (not shown) at this field that would indicate frozen moments or a strongly first-order phase transition, respectively. 

\subsubsection{Magnetic phases}

The presence of two magnetic phases in RuCl$_{3}$ is closely tied to structural details (see discussion in [5]) and thus their volume fraction can be expected to vary sensitively with stacking faults, strain, etc. In general we observe both magnetic transitions in Ru$_{1-x}$Ir$_{x}$Cl$_{3}$ crystals. However, differences in the relative prominence of the $T_{N1}$ and $T_{N2}$ transitions in observed in bulk measurements suggest that the volume fractions of these phases vary between samples. Fig. S7(b) shows susceptibility curves for three as-grown Ir-substituted crystals from the same batch, qualitatively showing (i) dominant ABC phase, (ii) comparable contributions, and (iii) dominant ABAB phase. That similarly doped samples (i) and (iii) exhibit nearly single-phase ABC and ABAB characteristics, respectively, indicates that Ir-substitution does not inevitably produce the ABAB-type phase (at least for low substitution levels, $x \leq 0.07$). The multiphase tendency in Ru$_{1-x}$Ir$_{x}$Cl$_{3}$ crystals is thus most likely a result of the modified growth process needed to incorporate Ir in transported crystals. 

\begin{figure}
\centering
\includegraphics[scale=0.35]{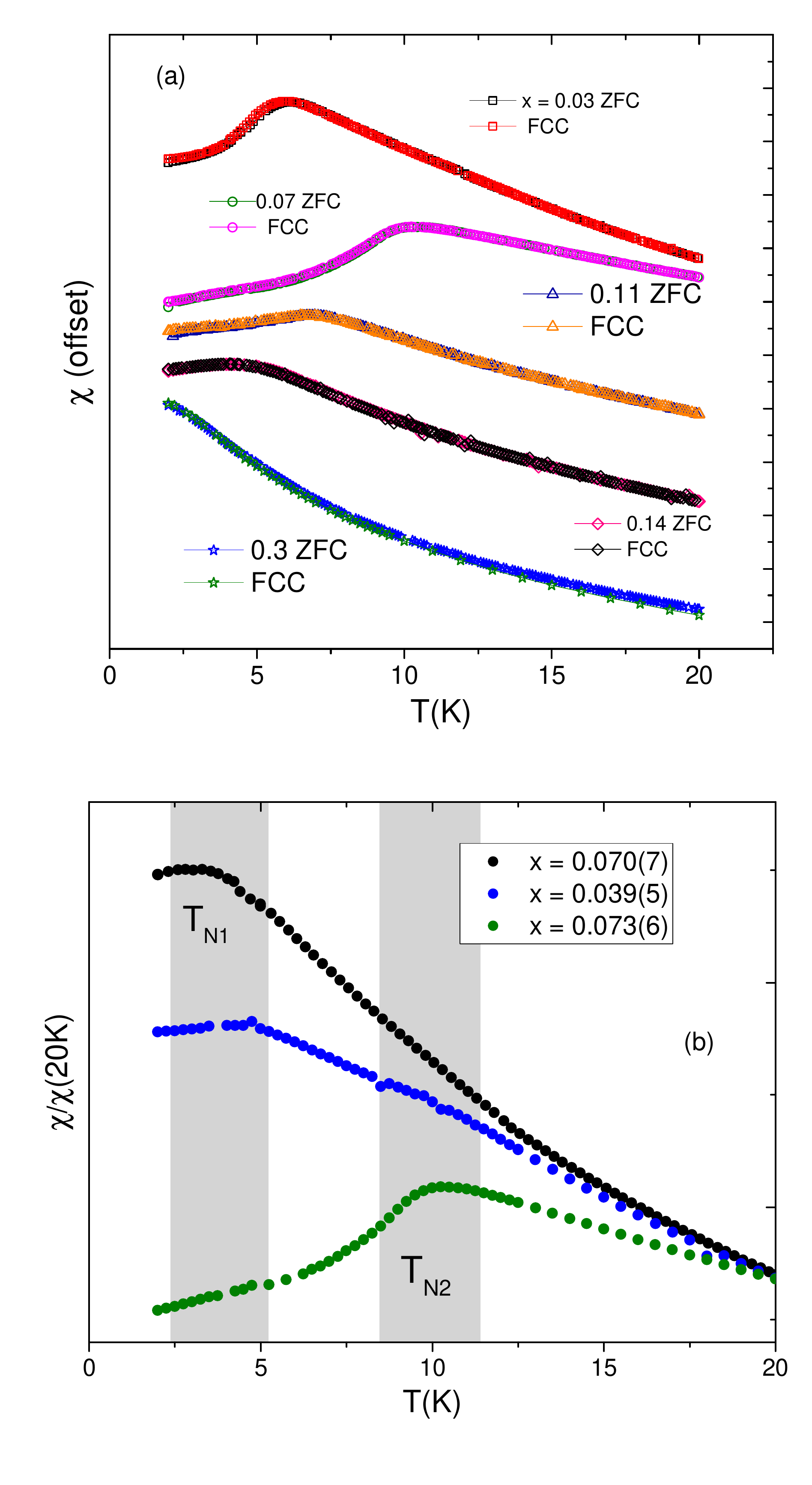}
\caption{(a) Susceptibility curves of Ru$_{1-x}$Ir$_{x}$Cl$_{3}$ single crystals collected under zero-field-cooled warming (ZFC) and field-cooled cooling (FC) protocols with a magnetic field of 1 T applied in the $ab$ plane. (b) Susceptibility curves, normalized to $\chi$ (20 K) for three Ir-substituted crystals from the same batch indicating a variation of the ABC and ABAB magnetic phase fraction between samples, as estimated from the relative prominence of $T_{N1}$ and $T_{N2}$.  } 
\end{figure}

\subsubsection{Field-dependent magnetization}

Field-dependent magnetization curves for Ru$_{1-x}$Ir$_{x}$Cl$_{3}$ single crystals at 2 K are shown in Fig. S8 over several substitution ranges. Magnetization initially decreases with $x$ up until $x \sim 0.1$  and the $M(B)$ curves retain an upward curvature (Fig. S8a). In the range $0.09 < x < 0.14$ magnetization increases and a switch in the concavity of $M(B)$ occurs, with an approximately linear field dependence in the $x = 0.11$ sample (Fig. S8b). At higher $x$, $M(B)$ is concave down and magnetization decreases continuously (Fig. S8c). 

We note that in the parent compound RuCl$_{3}$, field-dependent magnetization curves below $T_{N}$ switch from concave up to concave down around the critical field for suppression of the zigzag magnetic order $B_{C} \sim 7.5$~T, as shown in the inset of Fig. S8(a). It may be speculated that the Ir-induced curvature change in the substituted samples is related to a lowering of this critical field, with $B_{C} \rightarrow 0$ for $x \rightarrow 0.1$. However, no sharp kink is observed in the derivative $dM/dB$ for the substituted samples (Inset, Fig. S8b) up to the maximum field of 7 T available, so that no statement can be from these data made on the possible dilution-dependence of the critical field. 

\begin{figure*}
\centering
\includegraphics[width=1\linewidth]{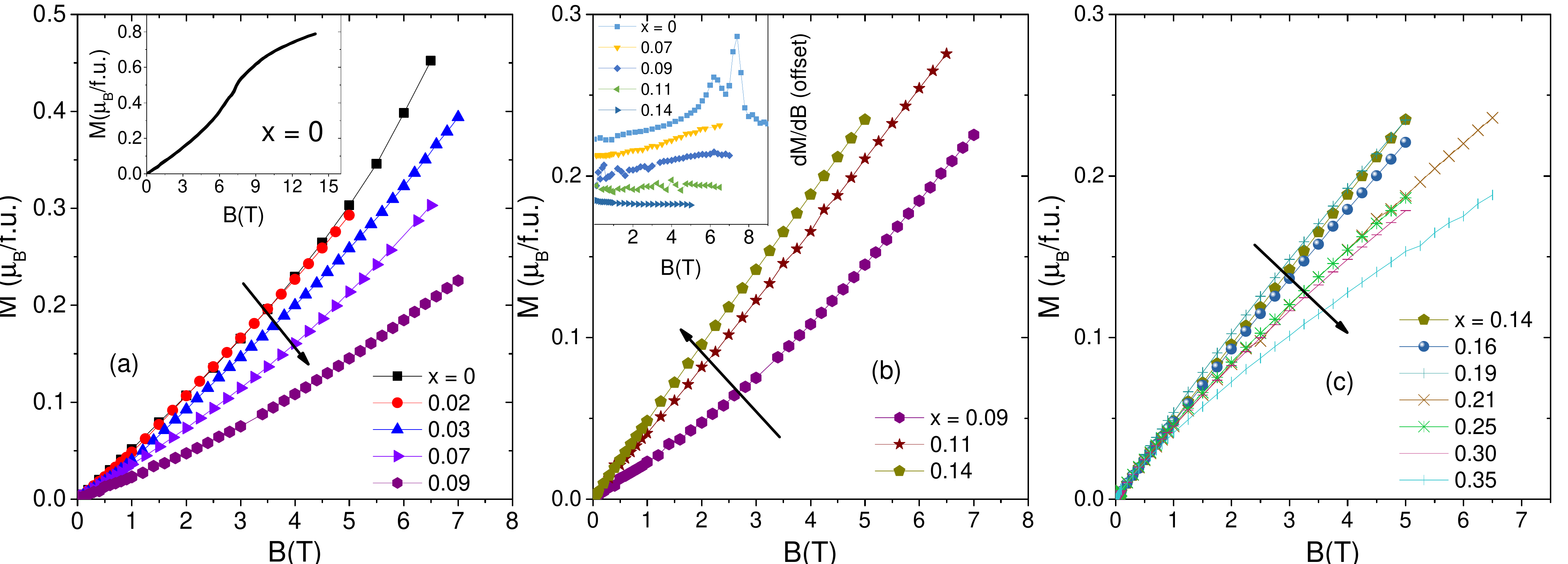}
\caption{Magnetization curves at 2 K with the magnetic field applied in the ab plane for Ru$_{1-x}$Ir$_{x}$Cl$_{3}$ crystals with (a) $0 \leq x \leq 0.09$, (b) $0.09 \leq x \leq 0.14$, and (c) $0.14 \leq x \leq 0.35$. Data were collected on two SQUID instruments with maximum magnetic fields of 5 T and 7 T. Inset (a): high-field dc magnetization curve for $x = 0$ measured on a 14 T PPMS with an ACMS insert. Inset (b): First derivative $dM/dB$ of magnetization vs. field for several compositions.} 
\end{figure*}

\begin{figure*}
\centering
\includegraphics[scale=1.5]{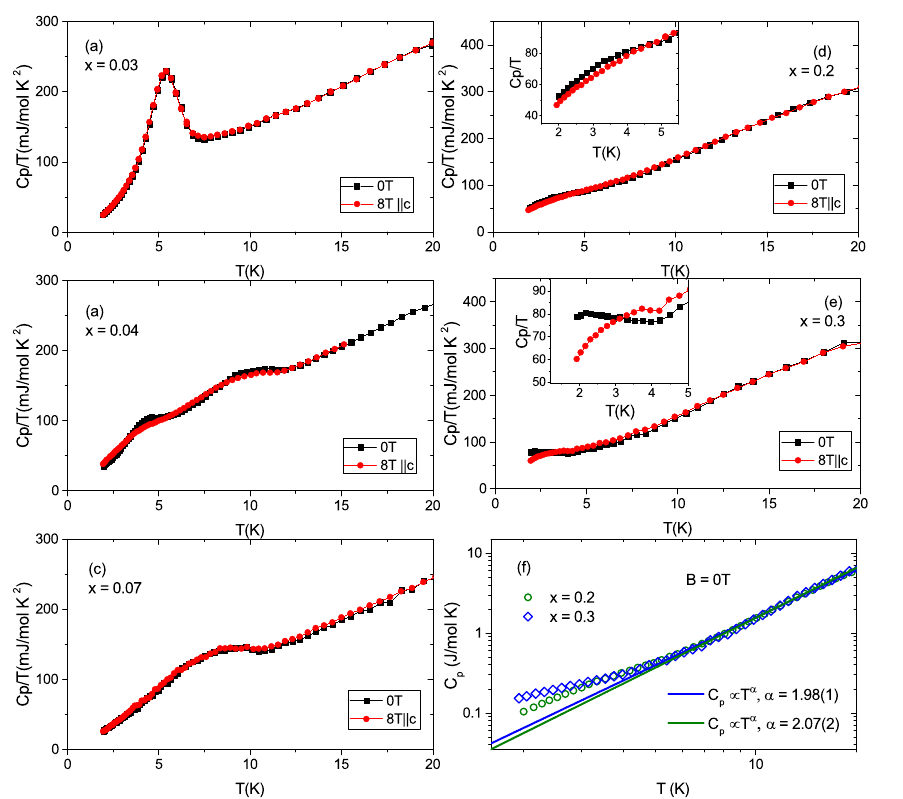}
\caption{Heat capacity curves $C_{p} / T$ vs. $T$ in zero field and with an 8 T field applied perpendicular to the honeycomb plane in Ru$_{1-x}$Ir$_{x}$Cl$_{3}$ crystals with (a) $x = 0.03$, (b) $x = 0.04$, (c) $x = 0.07$, (d) $x = 0.2$, and (e) $x = 0.3$. Insets in (d) and (e) are magnifications of the low-temperature region. (f) $C_{p}$ vs. $T$ for high Ir concentration, $x = 0.2$ and $x = 0.3$, showing a departure from $C_{p} \propto T^{2}$ behavior below $\sim$ 6 K where curves exhibit weak field dependence. } 
\end{figure*} 

\subsubsection{Heat capacity with $B \neq 0$}

The anomalies at $T_{N1}$ and $T_{N2}$ in heat capacity curves were not affected by a magnetic field applied perpendicular to the $ab$ plane (the only configuration available in our experimental setup) for small $x$ (Fig. S9a-c), in agreement with literature on the parent compound [4]. At larger substitution levels $x = 0.2$ and 0.3, where susceptibility measurements indicate the magnetic transitions are suppressed below 2 K, an upturn is visible in $C_{p} / T$ vs. $T$ curves approaching the base temperature of 2 K (Fig. S9d-e), and $C_{p}$ departs from a nearly $T^{2}$ behavior below $\sim$ 5 K (Fig. S9f). Magnetic field partially suppresses this upturn (insets, Fig. S9d,e), indicating that it has a magnetic origin. The feature may be related to short-range magnetic correlations, or represent the start of a very broad transition centered below 2 K. The emergence of weak field dependence with $B~||~ c$ points to a possible change in the magnetic anisotropies of the parent compound at large $x$, however a more detailed study with variation of the applied field direction in the crystals is required to explore this point.

\section{References}

[1] K. Brodersen, F. Moers, and H. Schnering, Naturwissenschaften 52, 205 (1965). 

[2] P. D. d. Rotier and A. Yaouanc, J. Phys.: Condens. Matter 9, 9113 (1997).

[3] F. Lang, P. J. Baker, A. A. Haghighirad, Y. Li, D. Prabhakaran, R. Valenti, and S. J. Blundell, Phys. Rev. B 94, 020407 (2016).

[4] M. Majumder, M. Schmidt, H. Rosner, A. A. Tsirlin, H. Yasuoka, and M. Baenitz, Phys. Rev. B 91, 180401 (2015).

[5] H. B. Cao, A. Banerjee, J.-Q. Yan, C. A. Bridges, M. D. Lumsden, D. G. Mandrus, D. A. Tennant, B. C. Chakoumakos, and S. E. Nagler, Phys. Rev. B 93 (2016).

\end{document}